\documentclass[12pt]{book}
\usepackage{plenum}
\begin{document}
\chapter{Survival Probabilities of Disoriented Chiral Domains in 
Relativistic Heavy Ion Collisions}

\author{Rene Bellwied \refnote{1}, Sean Gavin \refnote{2}, and Tom Humanic
\refnote{3}}

\affiliation{\affnote{1}Wayne State University, Physics Department, Detroit\\
\affnote{2}University of Arizona, Physics Department, Tucson\\
\affnote{3}The Ohio State University, Physics Department, Columbus\\}

\section{INTRODUCTION}

Disoriented chiral condensates (DCC) were recently proposed as potential
signatures for chiral symmetry restoration [\cite{bjorken},\cite{anselm},
\cite{blaizot1},\cite{raja2}]. In the theory of DCC formation, the explicit 
chiral symmetry breaking which occurs during the phase transition from
a plasma phase, where all masses are zero, to normal nuclear matter,
where particles have mass, is accompanied by the formation of extended
domains in which the chiral field is misaligned with respect to the
true vacuum direction.

This phenomenon has been explained in the context of the linear sigma
model. Here the isospin symmetry is not conserved during the occurance
of explicit chiral symmetry breaking, due to a slight tilt in the potential. 
This leads to to an oscillatory behavior of the long wavelength modes of the 
pion field in the sigma model, which subsequently causes the emission of 
pions of a specific isospin so that the field can regain its isospin
symmetric ground state. In Jorgen Randrup's contribution to this conference
[\cite{jorgen}] the theoretical aspects of the DCC formation are described
in more detail.

At the time of emission the content of a disoriented domain should behave
like a Bose condensate. Typical features of a Bose condensate,
sometimes referred to as pion laser, are the emission of very low momentum,
coherent pions and an increase in pion multiplicity in addition to the
already mentioned isospin asymmetry. These effects were described in detail 
by Pratt [\cite{pratt}].
Domain formation is a non-perturbative phenomenon which should occur during
any transition. The main question for experimental high energy physics is
whether the domains formed are large enough and sufficiently long-lived to
be detected.

The prime accessible signature of DCC formation is the ratio of neutral to
charged pions in a certain range of phase space, which should exhibit 
non-statistical isospin fluctuations. If one defines the parameter f as the 
ratio of number of neutral pions over the number of all pions emitted from 
the collision, then a standard isospin distribution will lead to a value 
sharply peaked around 0.33, whereas a DCC sample exhibits a probability 
function described by P(f) = 1/2$\sqrt{f}$. The determination of f requires 
a good measurement of the number of neutral pions in a slice of phase space.
Although these measurements are known to be very difficult, several experiments,
in particular WA98 in the CERN Heavy Ion Beam[\cite{WA98}] and MINIMAX in the 
FERMILAB Proton-Anti-Proton Beam[\cite{MINIMAX}], have attempted to discover 
DDC formation on the basis of the isospin ratio. Until now there were no 
positive results to corroborate the original CENTAURO events, measured two 
decades ago in high energy cosmic ray collisions [\cite{centauro}]. 
The MINIMAX analysis, although not yet successful in proving the existence of 
DCC's, lead to the generation of independent cut parameters to reduce the 
background in the data sample.

In James Symons' contribution [\cite{symons}] to this conference presents
the possibility of measuring the number of neutral pions with the STAR detector 
at RHIC. Because of the large particle density at RHIC energies (around 1500 
pions per unit rapidity), measurements of DCC formation can be performed on 
an event-by-event basis.

The goal of our study was to determine the survival probabilities of signatures 
of domain formation in basic charged particle observables, namely the momentum 
space parameters (transverse momentum and rapidity) and the coordinate space 
parameters (emission angle in azimuth and in pseudo-rapidity). 
The underlying assumption is that, if the domains stay confined to a certain 
phase space bin, the emission of pions should lead to non-statistical 
structures in the charged pion distributions on an event-by-event basis. 
Therefore a multi-dimensional moment analysis should lead to unambiguous 
signatures. Generally, DCC domains should stay localized in coordinate space 
and in momentum space due to collective motion during the hadronization phase.
Still, the effects of final state interactions on any hadronic signature have
to be taken into account. By using a dynamical transport code we try to show 
whether chirally disoriented domains can survive the hadronization and 
rescattering phases in a central relativistic heavy ion collision. A successful
measurement of such domains after freeze-out will depend on the domain
size, its localization in coordinate space and the manifestation of the 
disorientation at low transverse momentum. 
We therefore investigated the measurement of the transverse momentum, 
the rapidity spectrum and the pseudo-rapidity spectrum of pions below 200 MeV/c.
To simulate a realistic detector configuration we chose the STAR setup for
RHIC, in particular the tracking detector configuration of TPC and SVT, which
is sensitive to very low transverse momentum charged particles.

\medskip

The complete simulation chain consists of:

\medskip

a.) generating a DCC domain.

b.) embedding the domain in the background of a standard Au-Au event at RHIC
(in our case we chose HIJING [\cite{HIJING}] events).

c.) propagating the domain through the hadronization via a dynamical
transport code.

\medskip

The domain generation depends strongly on the domain parameters, in
particular the domain size and the energy density in the domain. Those
two parameters will determine the number and the momentum spectrum of the
pions in the domain.

The propagation depends strongly on the relative hadronization time of
the domain in comparison to the background pions. 

The next section tries to give a theoretical justification for the parameters
we have chosen for our study.

\newpage

{\bf Domain Size:}

For a second order equilibrium transition we expect the domain size to be about 
the correlation length. The mass of the pion sets the maximum correlation 
length to be around 1/135 MeV/c$^{2}$. These domains will be too small to be 
detected.

If the equilibrium transition is first order, then in principle we can 
expect larger domains and Kapusta and Vischer showed that by using Bjorken
hydrodynamics and relativistic nucleation theory [\cite{kapusta1}] one can
obtain sizeable domains. Still, at present, the second order phase transition
seems to be more realistic, based on lattice gauge calculations. Generally
second order transitions will not lead to large domains if the plasma 
hadronizes slowly and always stays close to chemical and thermal equilibrium.

For second order transition non-equilibrium transitions, Rajagopal and Wilczek
suggested the quench scenario in which the plasma cools very 
rapidly [\cite{raja1,raja2}]. A non-equilibrium situation could be caused
by significant supercooling before plasma hadronization. During
the cooling long wavelength modes of the pionic field grow exponentially
(spinodal decomposition). Gavin et al.[{\cite{gavin1}] and Boyanovsky et al.
[{\cite{boya}] showed independently that this exponential growth has a natural 
slowing process which leads again to relatively small but probably detectable 
domains (r = 2-3 fm). Although the quench scenario is not very likely, in 
particular because the plasma cooling should be slow compared to the 
de-excitation of the chiral field, in heavy ion collisions the rapid 
longitudinal expansion during hadronization of the quark-gluon phase may lead 
to sizeable disoriented configurations of the vacuum.

To explore the role of the medium in domain formation, Gavin and
M\"uller 
studied the dynamical evolution of the condensate in the presence of a
nonequilibrium bath of quasiparticles [\cite{gavin2}].  The three-dimensional 
expansion of the heat bath changed the effective potential rapidly
enough to create a quench-like condition.  The size scale was
for domains was found to be somewhat larger, perhaps $\sim 3-6$~fm,
compared to fixed-geomerty quench simulations [\cite{gavin1}] and
one-dimensional expansion.  This result has been verified by simulations by 
Randrup [\cite{jorgen3}].

Due to the relative uncertainty between the various non-equilibrium scenarios
we chose a rather conservative domain size of 3 fm for most of the simulations.

It should be noted that based on the sigma model the average transverse 
momentum of the pions contained in the domain is radius dependent. 
The mean p$_{T}$ is proportional to 1/r, which leads to a mean transverse 
momentum of about 100 MeV/c for a 3 fm domain radius. Fig.~1 shows the 
dependence of the spectral shape on the domain radius.
The condensation clearly generates a strong low momentum enhancement, as was
also pointed out by Ornik et al.[\cite{ornik}].


\begin{figure}[tbp]
\vspace*{3.in}
\caption{\it DCC domain transverse momentum distribution as a function of domain
radius.}
\end{figure}

{\bf Domain Energy Density:}

If the domain formation proceeds slowly and therefore close to equilibrium, 
the available energy is given by the tilt of the potential rather than the 
actual potential maximum (top of the 'Mexican Hat'). 

In this case the energy density in a domain is defined to be

\begin{equation}
 2 m_\pi^2 f_\pi^2
\end{equation} 

based on the explicit symmetry breaking term in the linear sigma model. 
This leads to a $\Delta$V = 40 MeV/fm$^{3}$, which in return defines the number 
of pions in a domain to about

\begin{equation}
N = \Delta V \times 4\pi/3 \times r^3/ m_\pi = 4500/140 = 32 pions, 
\end{equation}

assuming a 3 fm domain radius.

For a measurable effect, in particular in the environment of a central heavy ion
collision, we have to assume that either larger domains are formed or that
the energy density is enhanced due to plasma formation. An enhanced
energy density is conceivable, in particular in the context of a sudden
quench. Here the domain will start from a chirally symmetric distribution,
which means it will start out from the highest point on the 'Mexican Hat'
distribution. Therefore the domain in the condensed phase will absorb
the full potential energy plus a kinetic energy contribution from the initial 
conditions which could easily raise the energy density to more than
100 MeV/fm$^{3}$.  

Fig.2 shows the dependence of the number of pions on the energy density
and the domain size. In the simulations we varied the energy density between 
reasonable minimum and the maximum values, based on the actual sigma model 
potential, to adjust the number of pions in the domain to a measurable level.


\begin{figure}[hh]
\vspace*{3.in}
\caption{\it Dependence of DCC strength (number of pions) on energy density 
and domain radius}
\end{figure}

\clearpage

{\bf Hadronization Time:}

The proper hadronization time for the pions generated in a central Au-Au
collisions is set to be 1 fm/c, assuming a Bjorken evolution scenario.
 
Based on the dynamical treatment of the domain evolution within the linear
$\sigma$ model, Randrup [\cite{jorgen}] and Rajagopal [\cite{private}] argue
that it requires some time for the wavelength modes to oscillate and generate
domains. In particular Randrup's contribution is very quantitative and
shows that the typical domain takes about 4 fm/c to hadronize. During
this time the domain will not undergo any final state interactions, which
affects the survival probability, because the number of interactions is 
particle density dependent, and the expansion of the hadronized background 
will proceed between 1 fm/c and 4 fm/c to a level at which the interaction 
probability is seriously reduced.
On the other hand it was pointed out by Kluger [\cite{kluger}]
that if one chooses a late proper hadronization time (e.g. $\tau$ $\geq$ 2 fm/c)
the probability of domain formation and in particular domain growth might
actually decrease accordingly.

For our simulations we set the hadronization time of domain and background
to the same value, namely 1 fm/c, which should be regarded as worst case 
scenario, based on the dynamical mean field simulations mentioned above.

\section{RESULTS}

The dynamic transport code employed in these calculations was written
by T. Humanic and is described in detail elsewhere [\cite{humanic}].
The code is a kinetic model chosen to describe the evolution of the system 
after hadronization. Rescattering is
simulated using a Monte Carlo cascade calculation which assumes strong 
binary collisions between hadrons. Besides more common hadrons such as pions, 
kaons, and nucleons, the calculation also includes the $\rho$, $\omega$, $\eta$, 
$\eta$', $\phi$, $\Delta$, and K$^{*}$-resonances. Resonances can be present at 
hadronization and also can be produced as a result of rescattering. 
Relativistic kinematics is used throughout. Isospin-averaged scattering 
cross sections are taken from Prakash et al.[\cite{prakash}].
The domain itself is embedded into the hadronic background at the proper
hadronization time. The subsequent transport code is a boost invariant model
which follows the expansion rules of a Bjorken inside-outside cascade.

Our main calculation is based on 150 charged pions in a single domain with an 
average transverse momentum of about 100 MeV/c. Based on Fig.~2, these 
conditions can 
be accomplished either through an energy density of 150 MeV/fm$^{3}$ at the 
minimal measurable domain size (3 fm) or a 6 fm domain at the minimal available 
energy density (30 MeV/fm$^{3}$). The initial domain size determines the initial 
pseudo-rapidity distribution, therefore the 3 fm domain is preferred to confine 
the domain to a small bin in configuration space.
Assuming that the domain is confined within the acceptance of the STAR detector 
tracking system ($\eta$=$\pm$1) the background contains around 1500 charged 
pions at the maximum efficiency. The unfavorable signal-to-noise ratio of 1:10,
can be improved by correlating the pseudo-rapidity distribution with the 
transverse momentum distribution, see Fig.8. Our initial intent, though, was 
to detect the domain by simply analyzing the pseudo-rapidity spectrum itself. 
Fig.3(a) shows the spectrum of all charged pions in the STAR detector
acceptance, assuming that the domain stays confined to r = 3 fm until 
freeze-out and does not undergo rescattering. 
Fig.3(b) shows the result of the transport code, including rescattering, 
applied to the same single Au-Au event. 
Obviously the low momentum pions scatter and the domain is diluted. Based on 
the statistical error in an event-by-event measurement (around 7$\%$), 
the domain can not be detected at this level without using more sophisticated 
analysis procedures.

\begin{figure}[tbp]
\vspace*{4.in}
\caption{\it DCC rapidity distribution with and without rescattering in a
central Au-Au heavy ion collision at RHIC (as simulated by HIJING)}
\end{figure}

We decided to apply a method, first suggested by Huang et al., which is based 
on multi-dimensional wavelet transformation. The method is described in detail 
elsewhere [\cite{wavelet}], but the main way of scanning the distribution is by 
assigning a wavelet function to represent the data. Wavelet functions are 
invertible and orthogonal and are used to represent spiky distribution, in 
contrast to Fourier transforms which are used to parameterize smooth deviations.
In this case the pseudo-rapidity distribution is described by a mother 
function (general distribution) and a father function (deviations from general 
distribution), which leads to a multi resolution moment analysis in which the 
coefficient of the father function is a measure of the size of the effect to 
be measured. 

\begin{equation}
f(\eta) = f^J(\eta) = \sum_{k=0}^{2^{j-1}-1} f_{j-1,k} \phi_{j-1,k}(\eta) +
\sum_{k=0}^{2^{j-1}-1} F_{j-1,k} \psi_{j-1,k}(\eta)
\end{equation}

f$_{j-1,k}$ is the mother function coefficient and F$_{j-1,k}$ is the
father function coefficient. The parameter j is determined by the
resolution scale, k denotes the position at each scale.

The wavelet equation used in this analysis has a Haar Basis. Whether another
Basis might be better suited to measure the effect will depend on further
studies. For other more general references regarding wavelet analysis we 
refer the reader to a paper by J. Randrup [\cite{jorgen2}] and the references 
therein.

The power spectrum of the father function coefficients is a measure for
the size of the deviation from a smooth distribution.

\begin{equation}
P_j = 1/2^j \sum_{k=0}^{2^j-1} \mid{F_{jk}}\mid^2
\end{equation}

The higher the power of the father function coefficients the larger the
deviation from the non-disturbed distribution. The pseudo-rapidity space is
scanned in defined step sizes which relate to the size of the deviation.
Even pure event generator events (e.g. HIJING central Au-Au) have
non zero coefficients due to the statistical fluctuations on an 
event-by-event basis. In addition, the detector resolution adds another
potential deviation to an ideal smooth distribution.

This analysis utilizes the size and the location of a domain simultaneously
for identification purposes. We apply the analysis on an event-by-event
basis which is slightly different form the original paper [\cite{wavelet}].
Here the deviation of the power spectrum is not an averaged effect on a
data sample but rather describes the deviation of a single cell from the
background as defined by the non-perturbed remainder of the pseudo-rapidity 
distribution. Therefore a very large domain (larger than the resolution 
scale = 2 units of $\eta$) or a complete condensation will not lead to a 
measurable effect with this method.

Fig.~4a shows the result of the wavelet analysis applied to the distributions
shown in Fig.~3 and to a standard single event HIJING distribution.
The wavelet analysis allows to determine a domain of size 3 fm and energy 
density about 100 MeV/fm$^{3}$ even after rescattering. The general x-scale, 
the step size of the binning, is translated into a pseudo-rapidity scale. 
After rescattering the domain covers around 0.5 units, whereas before 
rescattering the domain was confined to about 0.3 units.
Although the re-scattering weakens the domain it does not fully destroy its
main feature namely its confinement to a rather small bin in configuration
space. 

A more general simulation, which shows the dependence of the detectability of 
domains based on their respective size and energy density is shown in Fig.~4b. 
We conclude that at a nominal domain size of 3 fm, domains with an energy 
density as low as 50 MeV/fm$^{3}$ can be detected via a wavelet analysis even
after re-scattering.


\begin{figure}[tbp]
\vspace*{4.in}
\caption{\it a.)Multi-resolution analysis for the rapidity distributions shown in
Fig.~3 plus an undisturbed HIJING event.
b.) Variation in multi-resolution analysis as a function of
energy density in an r=3 fm DCC domain}
\end{figure}

An additional effect, though not included in this calculation,
that might contribute to the survival probability of the domain is Bose 
cascading.
This effect was first observed in condensed matter [\cite{bosecasc}], but it
seems to affect the heavy ion spectra as well. A recent evaluation of the
effect of Bose kinetics on the low momentum part of the pion spectrum
can be found in [\cite{gerd1}]. The authors conclude that the effect of
Bose enhanced scattering leads to a doubling of the pion
cross section for momenta below 100 MeV/c in central heavy ion collisions
at CERN fixed target energies. Therefore we might observe two competing
effects in the resulting spectrum. Low momentum domain pions
rescattering from higher momentum target matter gain momentum, but in parallel
part of the spectator matter is cascading to lower momenta.
Therefore the number of particles at very low momentum stays almost constant
and their rapidity distribution increases only slightly in width. A detailed
simulation of this effect is underway [\cite{gerdy}].

The amount of rescattering depends strongly on the location of the domain
in configuration space at the time of domain formation. In our study we place
the domain in the center of the fireball at the proper hadronization time.
Fig.~5 displays
the pseudo-rapidity distribution of DCC pions 
for a standard domain (r = 3 fm, $\epsilon$ = 100 MeV/fm$^{3}$)
at freeze-out as a function
of the relative distance of the domain center from the surface of the
fireball in configuration space at the time of domain hadronization. 
The domain undergoes lesser re-scattering the further inside
the fireball it is created, which subsequently leads to a more confined 
pseudo-rapidity distribution for domains produced in the core of the fireball. 
This is due the kinematics of the domain particles with respect to the
fireball particles. The low momentum domain pions expand slower than the 
fireball pions, and therefore decouple early from the fireball. A domain at 
the fringes of the fireball, though, will experience the most final state 
interactions, because the radially expanding fireball will traverse through 
the domain before decoupling and freeze-out.

\begin{figure}[tbp]
\vspace*{4.in}
\caption{\it Comparison of rapidity distributions of DCC pions as a function of
the relative distance in configuration space between the domain center and the
surface of the fireball.}
\end{figure}

A recent study by V.Koch
[\cite{gerdy}] shows that in his transport model domains can not survive
the rescattering. In these instances, the domains were placed at the surface
of the fireball in the rest frame of the fireball. Based on our studies we
conclude that a domain in the center of the fireball will freeze out first
and therefore experience the fewest final state interaction. We therefore 
believe that the results from Koch and also results presented by the RQMD
group at Quark Matter 97 are not necessarily incompatible 
with our simulations.

To further investigate the domain kinematics we compare in Fig.~6
the freeze-out time distribution of the pions in the hadronic
background and the domain pions. On average
the low momentum pions freeze-out faster, which at first seems counterintuitive.
The slower particles should have more interactions for they stay longer
in the interaction zone. This effect is reversed though for the very slowest
particles. In this instance the fireball traverses through the pions which
remain almost at rest. Therefore, the 
interaction zone decouples after a few tens of fm/c from the
pions of interest. This might be one of the reasons that leads to a 
high survival probability for DCC's throughout the re-scattering phase.

\begin{figure}[tbp]
\vspace*{4.in}
\caption{\it Freeze-out time for various components of the pion spectrum}
\end{figure}

It is worthwhile pointing out that an increase in the hadronization time
for the domain itself will certainly lead to an increased survival probability.
Assuming that the simulations by Rajagopal and Randrup are correct we can 
expect
hadronization times as large as 5 fm/c. From the freeze-out time spectrum
(Fig.~6) we deduce that the DCC pions have a freeze-out time of around
10 fm/c. Most of the final state interactions will occur in the
very early part of the fireball hadronization, which means that late
hadronization of the domains will lead to very large survival probabilities.
In addition the domain freeze-out time will depend on the spatial position 
of the domain with respect to the fireball (see Fig.~6). 

Fig.7 shows the effect of the rescattered domain on the transverse
momentum spectrum. Fig.7(a) shows the transverse momentum spectrum of
positively charged pions in the pseudo-rapidity range from -1 to +1
for a single central Au-Au collision. About 900 pions of a single isospin
are found within two units of pseudorapidity. Fig.7(b) shows the same event
including our standard DDC domain at freeze-out, after rescattering. The 
original number of 150 pions stays confined to the low momentum region. 
The ratio of "DCC" to "standard" pions below p$_{T}$=150 MeV/c is about 2:1, 
an enhancement which should be detectable with any tracking detector with 
modest tracking efficiency below 100 MeV/c. The mean of the distribution shifts 
from about 330 MeV/c to 300 MeV/c. A shift in the mean could be evaluated 
sufficiently fast to serve as a trigger signal for DCC detection.
The strong confinement in momentum space can also be used as an additional 
constraint on the multi-dimensional analysis of the pseudo-rapidity spectrum.

\begin{figure}[tbp]
\vspace*{4.in}
\caption{\it Comparison of the transverse momentum spectrum of positively
charged pions in the two central units of pseudo-rapidity with and without
a standard DCC domain}
\end{figure}

It should be pointed out that our simulations are based on the formation
of a single domain at the center of the fireball. In a sense the formation
of a single domain is highly unlikely. If the effect of chiral disorientation
occurs one would expect many domains to be formed. As soon as the domain
separation decreases below the size of the scattered domain as simulated
in our code, the effect in the pseudo-rapidity spectrum will be lost due to 
domain merging. This effect is well described by the Central Limit Theorem.
On the other hand each single domain will contribute to the low momentum
enhancement in the transverse momentum spectrum. Therefore the production
of multiple domains should lead to a stronger shift of the mean p$_{T}$
and to a stronger shape change of the spectrum towards lower p$_{T}$.

In Fig.~8 we show two-dimensional
distributions based on a simulation with five equally sized domains in a
central Au-Au HIJING event. The figure tries to qualitatively compare the
configuration space and momentum space distributions of pions after 10 fm/c
for an event with domains (a+c) and an event without domain (b+d). It seems
possible that by correlating the information from the spatial measurements
(azimuth and pseudo-rapidity distributions) with the information from the
momentum measurements one might be able to characterize events independent
of the number of domains per event.

\begin{figure}[tbp]
\vspace*{6.in}
\caption{\it Distribution of a HIJING event with and without domains
in configuration (a+c) and momentum space (b+d). The momentum space exhibits
strong low momentum components correlated with large fluctuations in
configuration space.}
\end{figure}

\clearpage

\section{CONCLUSIONS}

Based on a realistic dynamic transport code applied to relativistic heavy ion
collisions which contain domains of disoriented chiral condensate and/or
simple Bose condensates we conclude that low momentum particle domains can
survive hadronic final state interactions. Their detectability will depend
on the number of pions in the domain which in return depends on the domain
size and the energy density inside the domain. For the detector resolution
applicable to the STAR detector at RHIC we conclude that by applying a
multi-resolution analysis 3 fm radius domains can be detected on an event by
event basis if the energy density inside the domain exceeds 50 MeV/fm$^{3}$.
If the domain size increases, the energy density threshold can be lowered
accordingly. The formation of DCC's affects two event-by-event observables,
namely the pseudo-rapidity and transverse momentum distributions of the charged
pions. Multi domain generation in a single event might lead to a reduction
of the effect in the pseudo-rapidity distribution, but it will lead to an 
enhancement of the effect in the momentum distribution. Multi-dimensional 
correlation studies which employ cuts in momentum and in configuration space 
should therefore lead to conclusive results concerning the existence of 
disoriented chiral domains, even without measuring the isospin dependence of 
the domain pions.

Therefore we strongly suggest that RHIC experiments put special emphasis on
their low momentum coverage over a large range of pseudo-rapidity. Suggestions
to use a topology trigger to trigger on non-statistical fluctuations in
the rapidity distributions of charged particles will help to further select
potential DCC events. At this point, both PHOBOS and STAR seem to be capable
of measuring the effects of DCC formation in charged particle spectra.




\begin{numbibliography}
\bibitem{bjorken} J.D. Bjorken, {\em Inter. J. Mod. Phys.} {\bf A7} (1992) 4189
\bibitem{anselm} A.A. Anselm and M.G. Ryskin, {\em Phys. Lett.} {\bf B266} (1991) 482
\bibitem{blaizot1} J.P. Blaizot and A. Krzywicki, {\em Phys. Rev.} {\bf D46} (1992) 246
\bibitem{raja2} K. Rajagopal and F. Wilczek, {\em Nucl. Phys.} {\bf B404} (1993) 577
\bibitem{jorgen} J. Randrup, Contribution to these Proceedings (1998)
\bibitem{pratt} S. Pratt, {\em Phys. Lett.} {\bf B301} (1993) 159.
\bibitem{WA98} T. Nayak, Quark Matter 97, Tsukuba, Japan, to be published 
\bibitem{MINIMAX} T.C. Brooks et al., {\em Phys. Rev.} {\bf D55} (1997) 5667
\bibitem{centauro} C.M.G. Lattes, Y. Fujimoto, and S. Hasegawa, {\em Phys. Rep.} {\bf 65} (1980) 151
\bibitem{symons} J. Symons, Contribution to these Proceedings (1998)
\bibitem{HIJING} X.N. Wang and M. Gyulassy, {\bf Phys. Rev.} {\bf D44} (1991) 3501 
\bibitem{kapusta1} J.I. Kapusta and A.M. Srivastava, {\em Phys. Rev.} {\bf D50} (1994) 5379
\bibitem{raja1} K. Rajagopal and F. Wilczek, {\em Nucl. Phys.} {\bf B379} (1993) 395
\bibitem{gavin1} S. Gavin, A. Goksch and R.D. Pisarski. {\em Phys. Rev. Lett.} {\bf 72} (1994) 2143
\bibitem{boya} D. Boyanovsky, D.L. Lee {\em Phys. Rev.} {\bf D48} (1993) 800
\bibitem{gavin2} S. Gavin and B. Mueller, {\em Phys. Lett.} {\bf B329} (1994) 486
\bibitem{jorgen3} J. Randrup, {\em Nucl.Phys.} {\bf A616} (1996) 531
\bibitem{ornik} U. Ornik et al., Los Alamos Preprint (LA-UR-96-1615) and nucl-th/9808027 (1996)
\bibitem{kluger} Y. Kluger, Los Alamos Preprint (LA-UR-94-1566) and hep-ph-9405279 (1994)
\bibitem{private} K. Rajagopal, Presentation at the STAR Collaboration Meeting,
August 97
\bibitem{humanic} T.J. Humanic, {\em Phys. Rev.} {\bf C50} (1994) 2525 
\bibitem{prakash} M. Prakash, M. Prakash, R. Venugopalan, and G. Welke, {\em Phys. Rep.} {\bf 227} (1993) 321
\bibitem{wavelet} Z. Huang et al., {\em Phys. Rev.} {\bf D54} (1996) 750
\bibitem{jorgen2} J. Randrup, {\em Phys. Rev.} {\bf D56} (1997) 4392
\bibitem{bosecasc} Y.B. Zeldovich and E.V. Levich, {\em Sov. Phys. JETP} {\bf 28} (1969) 1287
\bibitem{gerd1} G. Welke, and G. Bertsch, {\em Phys. Rev.} {\bf C45}
(1992) 1403.
\bibitem{gerdy} V. Koch and G. Welke, private communication (1998)
\end{numbibliography}
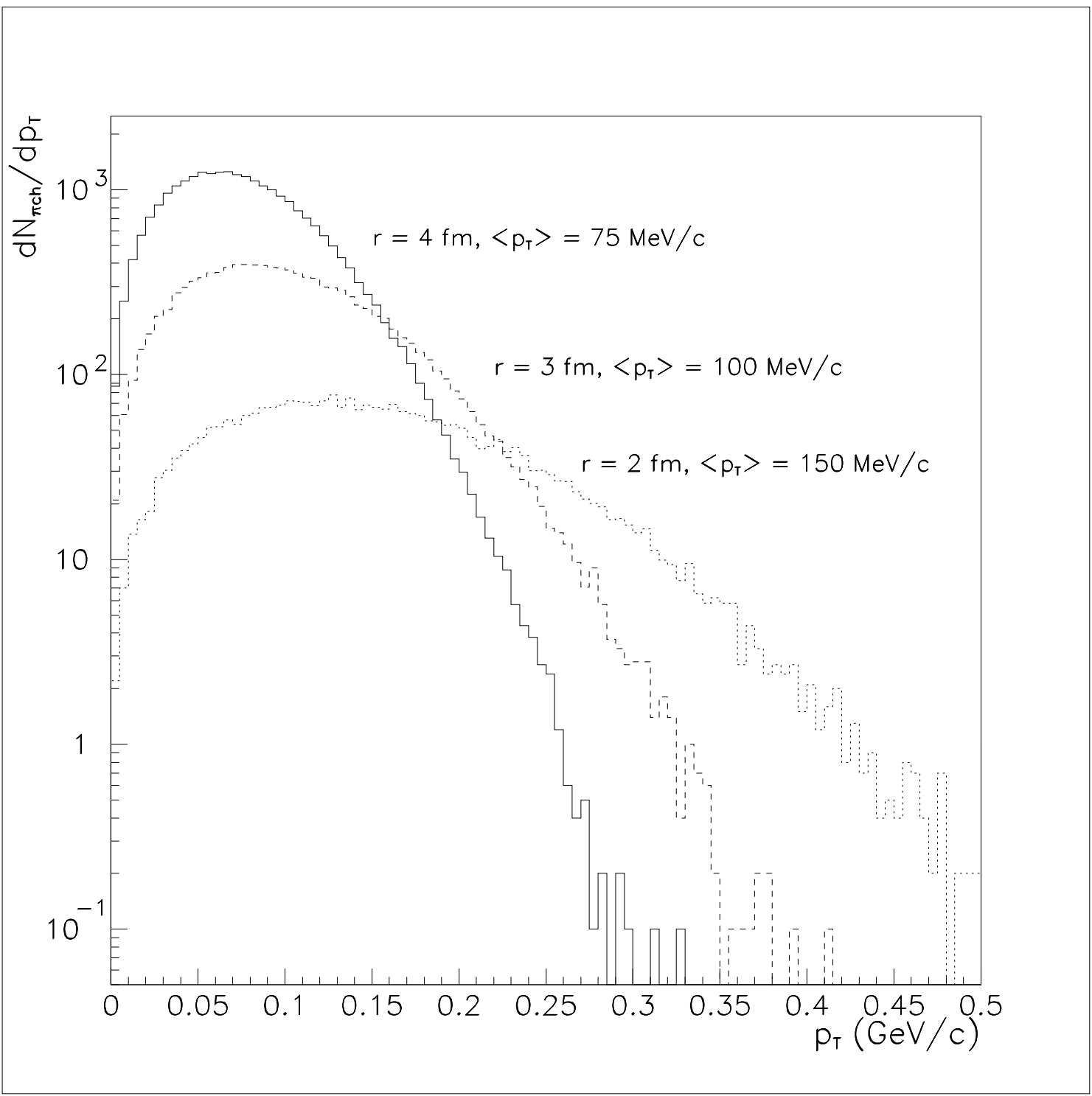
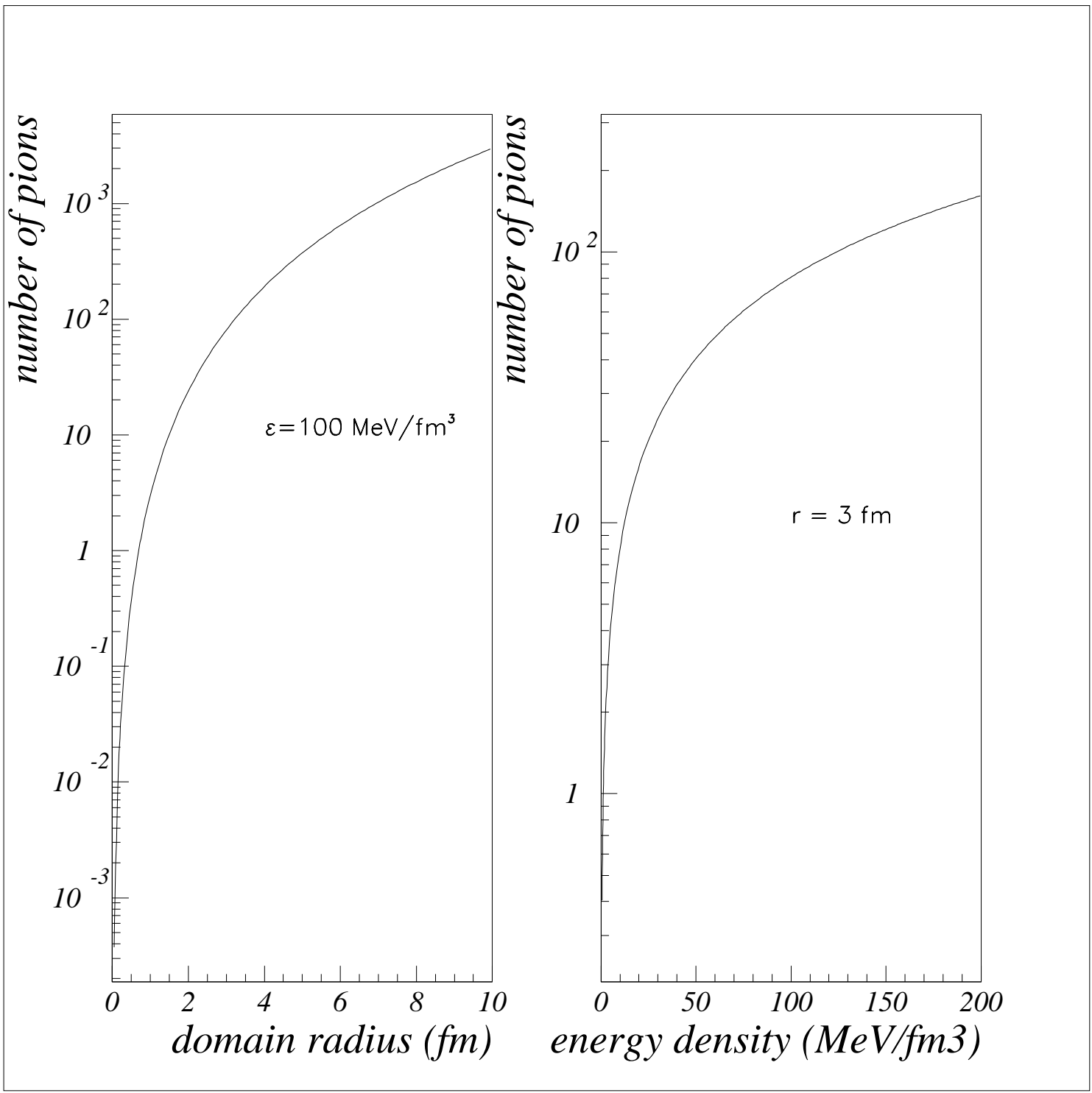
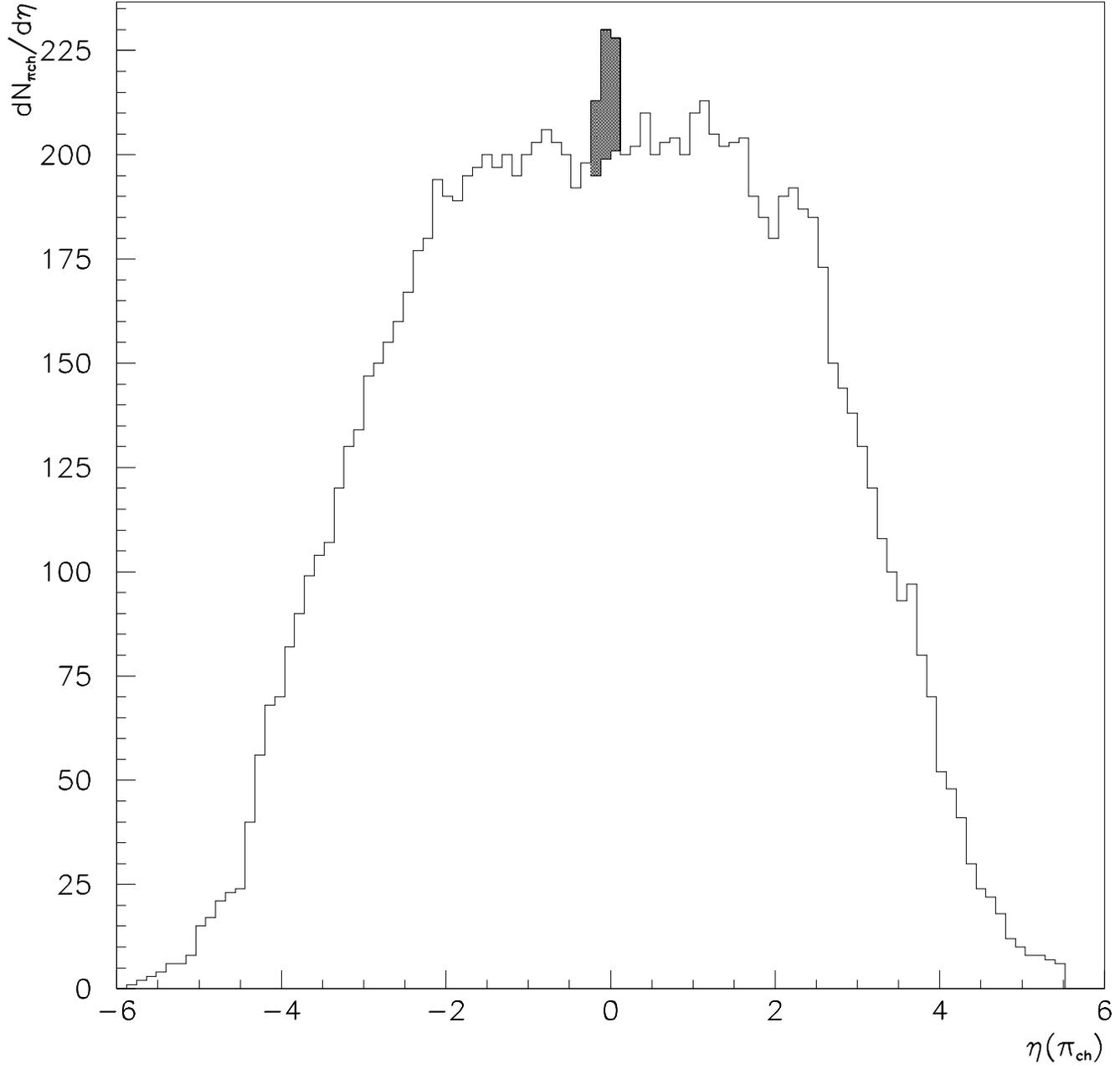
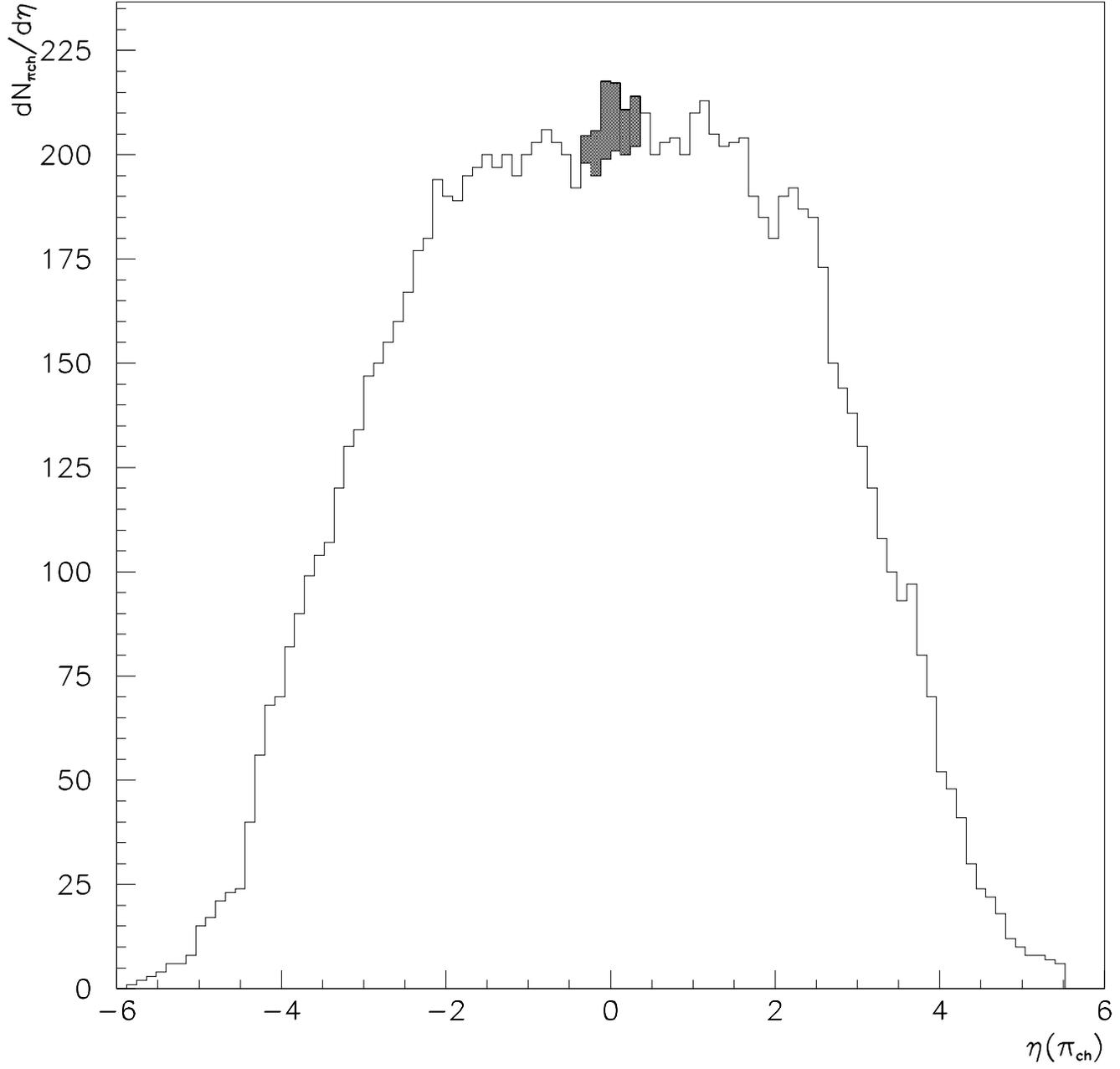
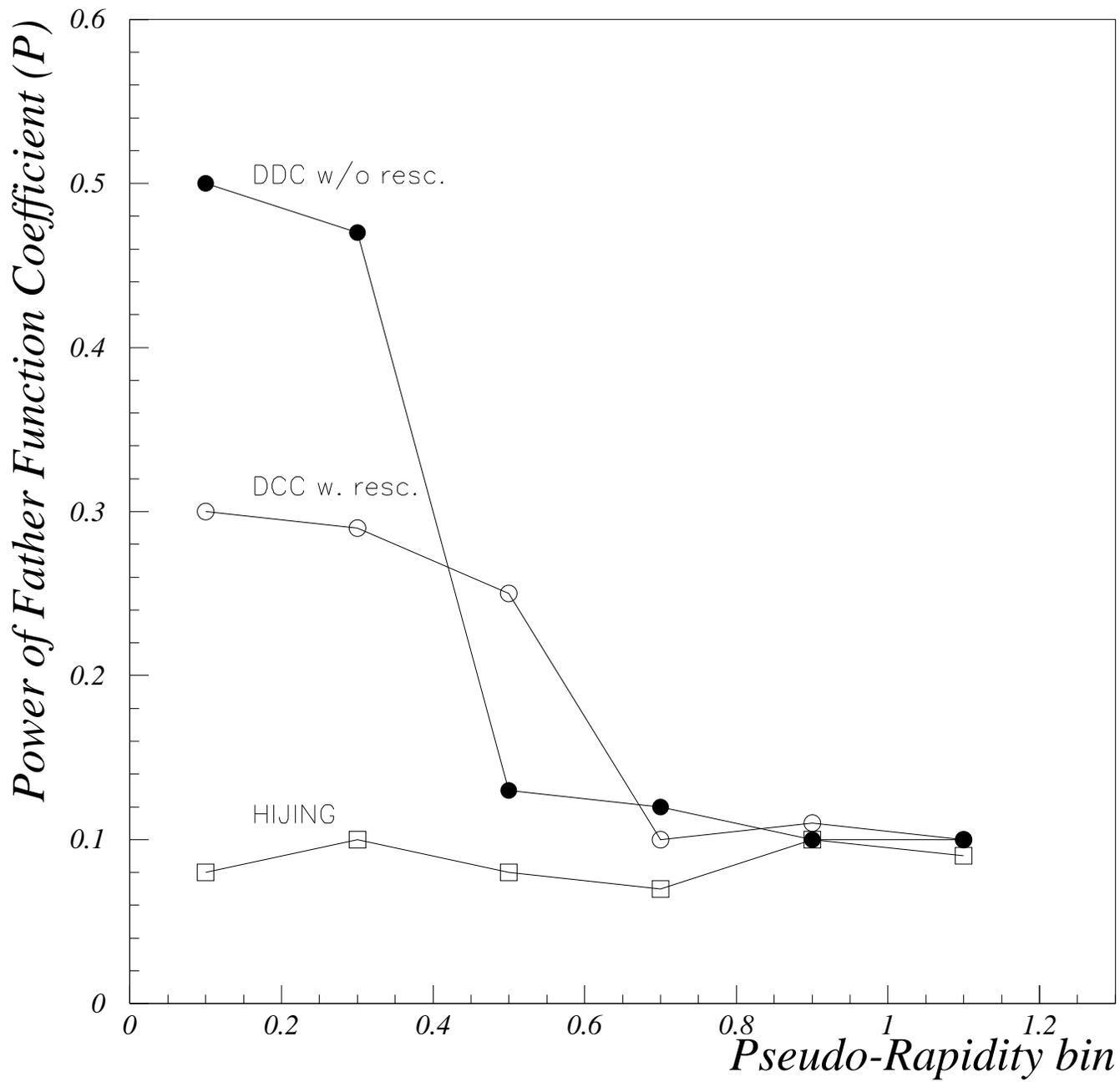
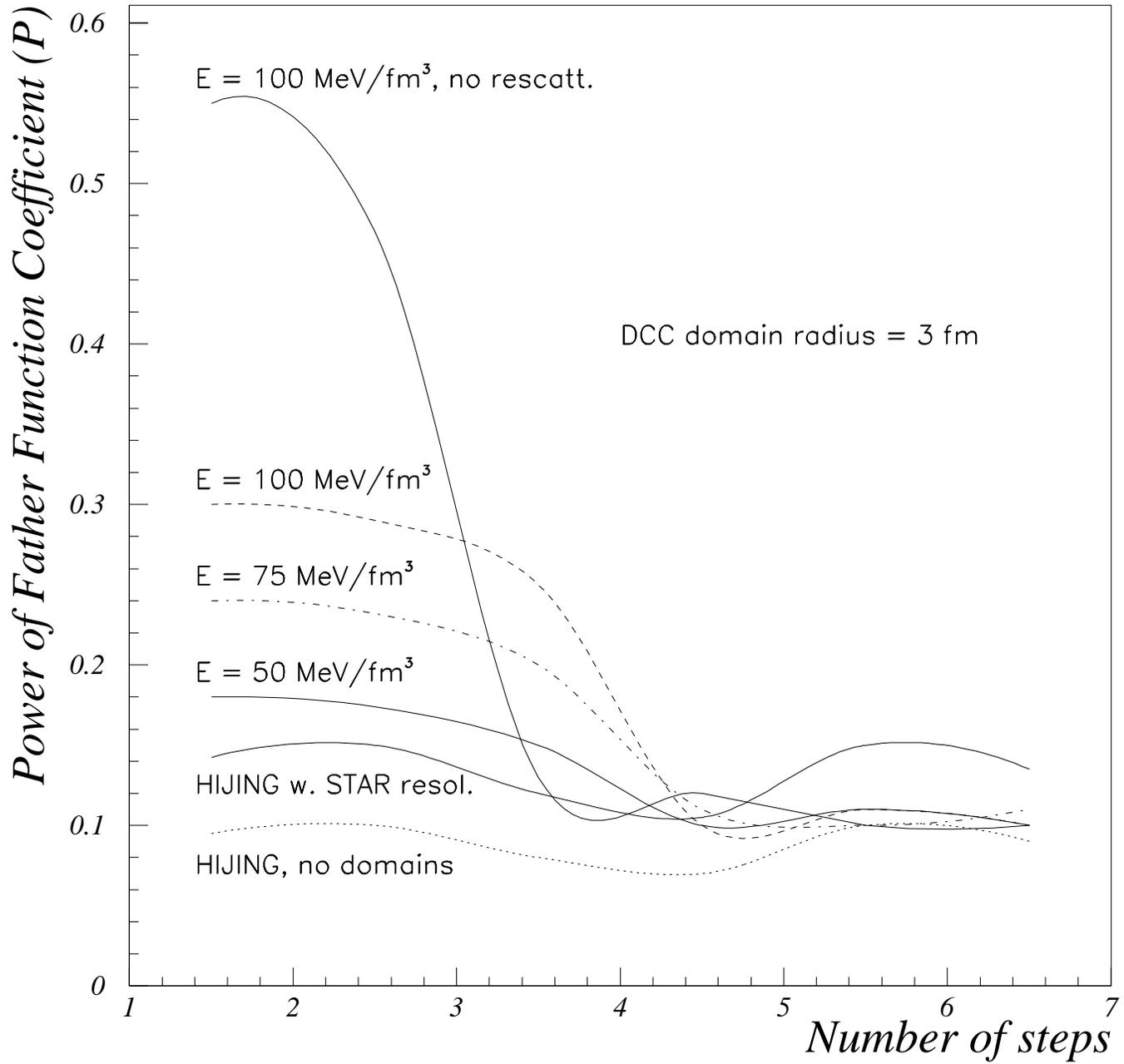
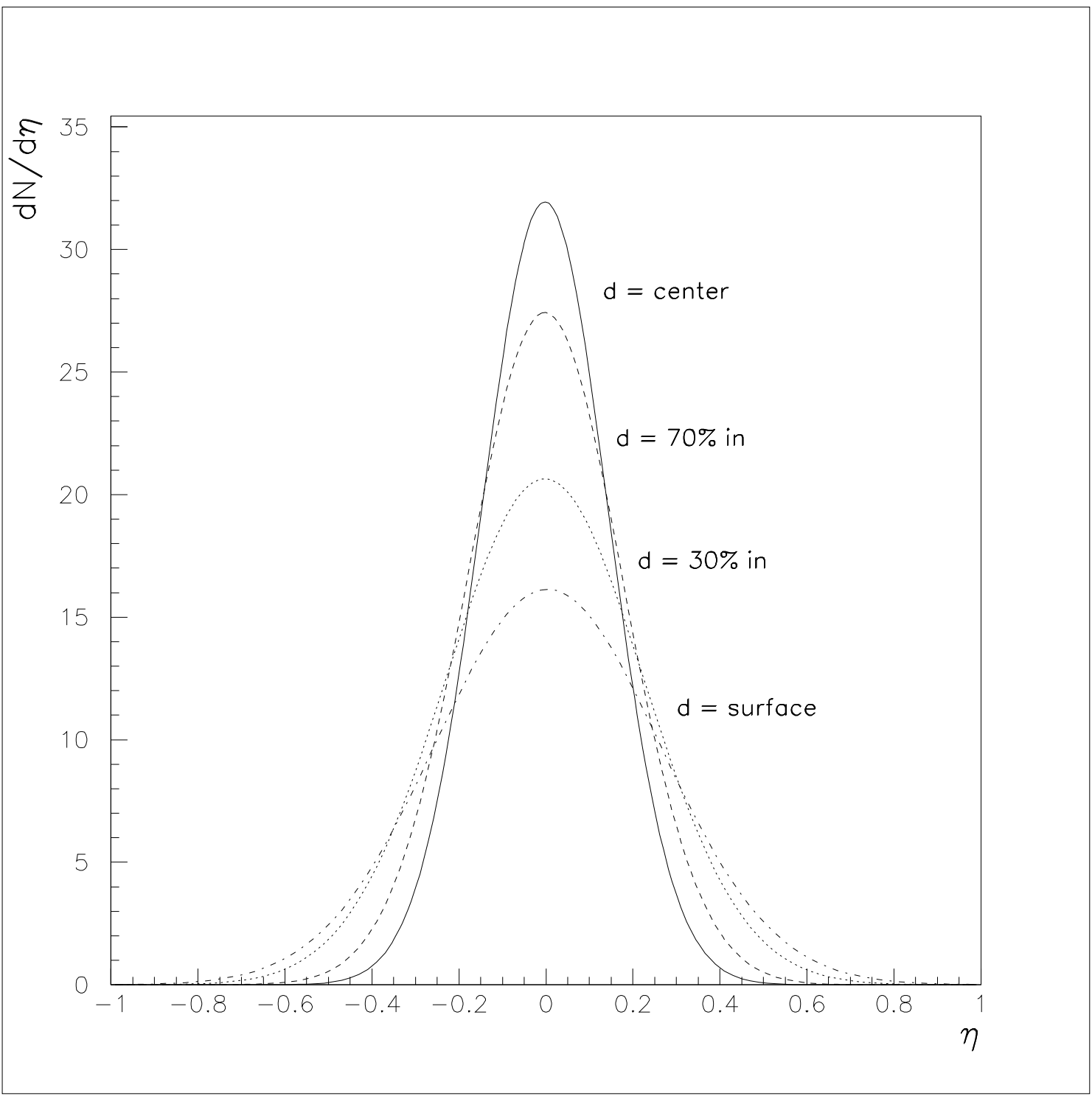
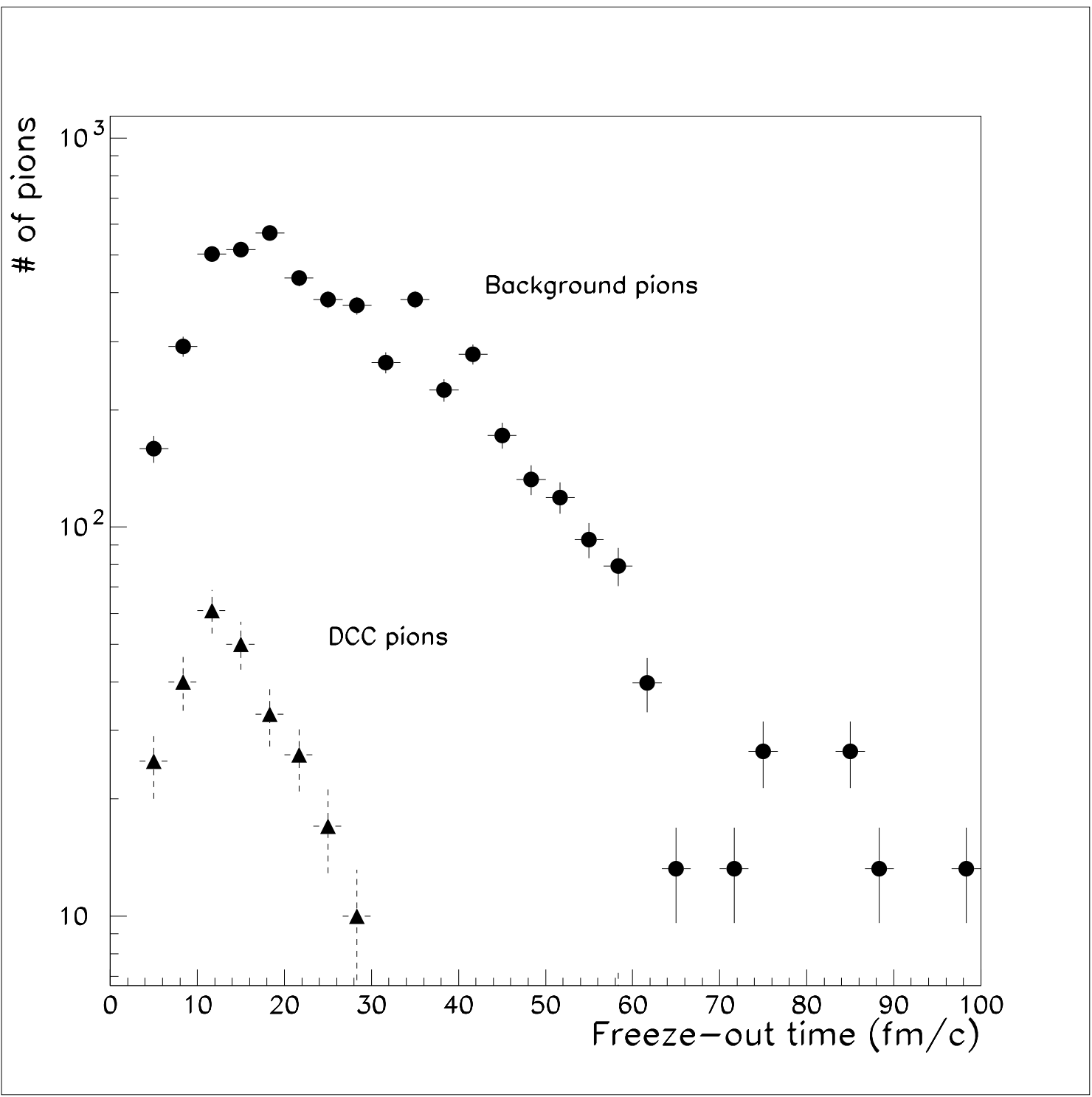
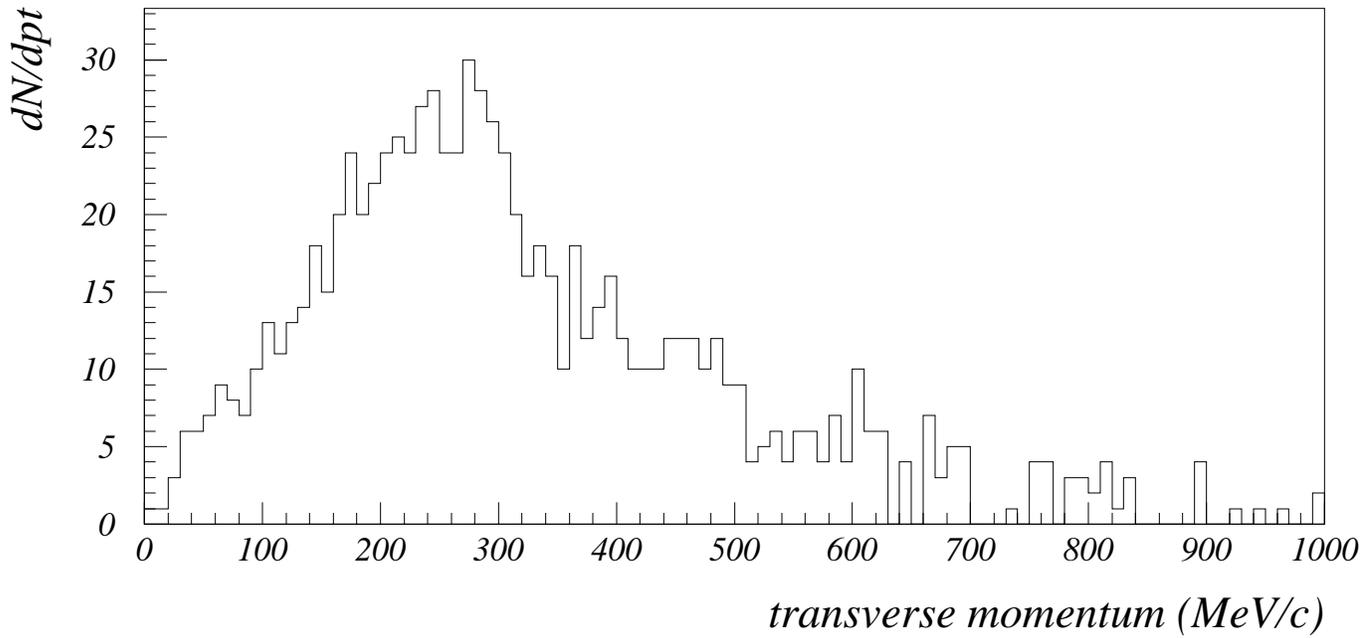
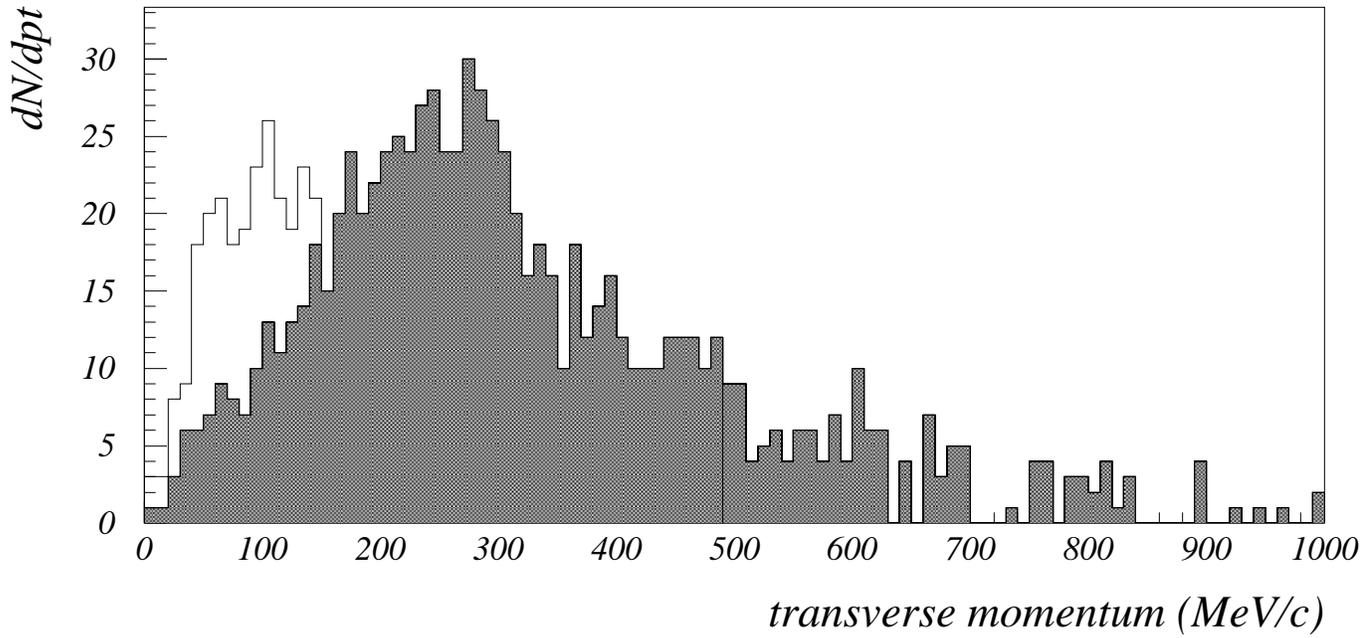
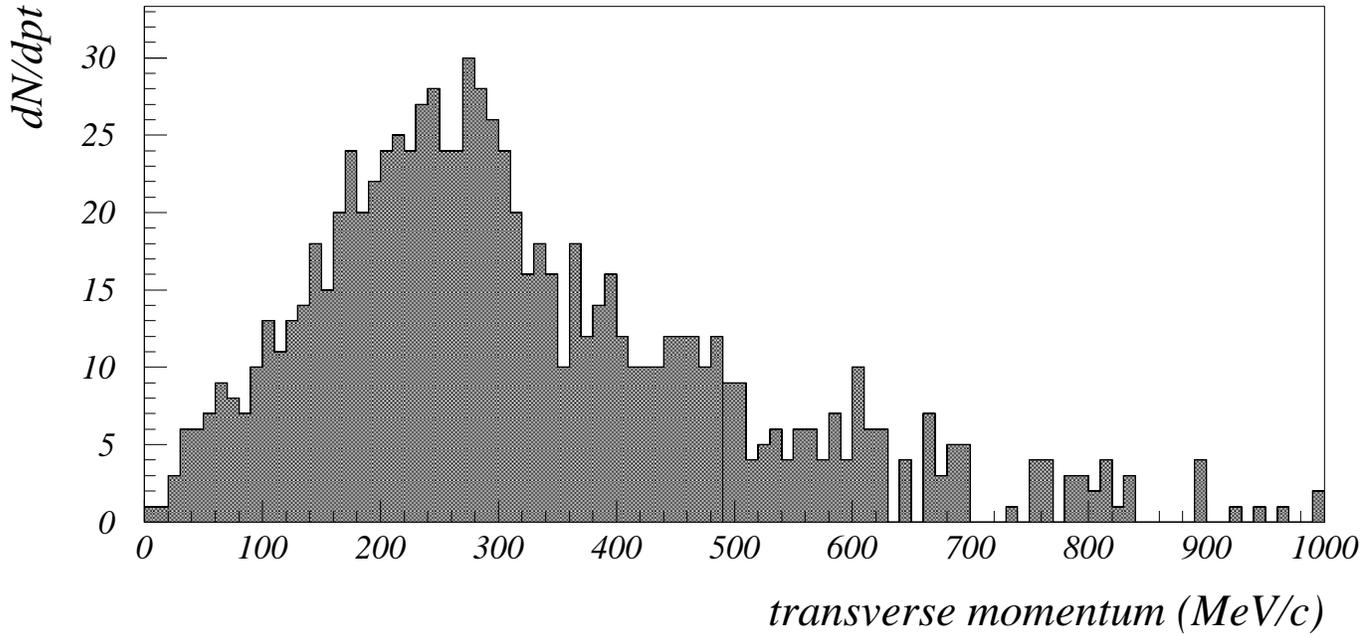
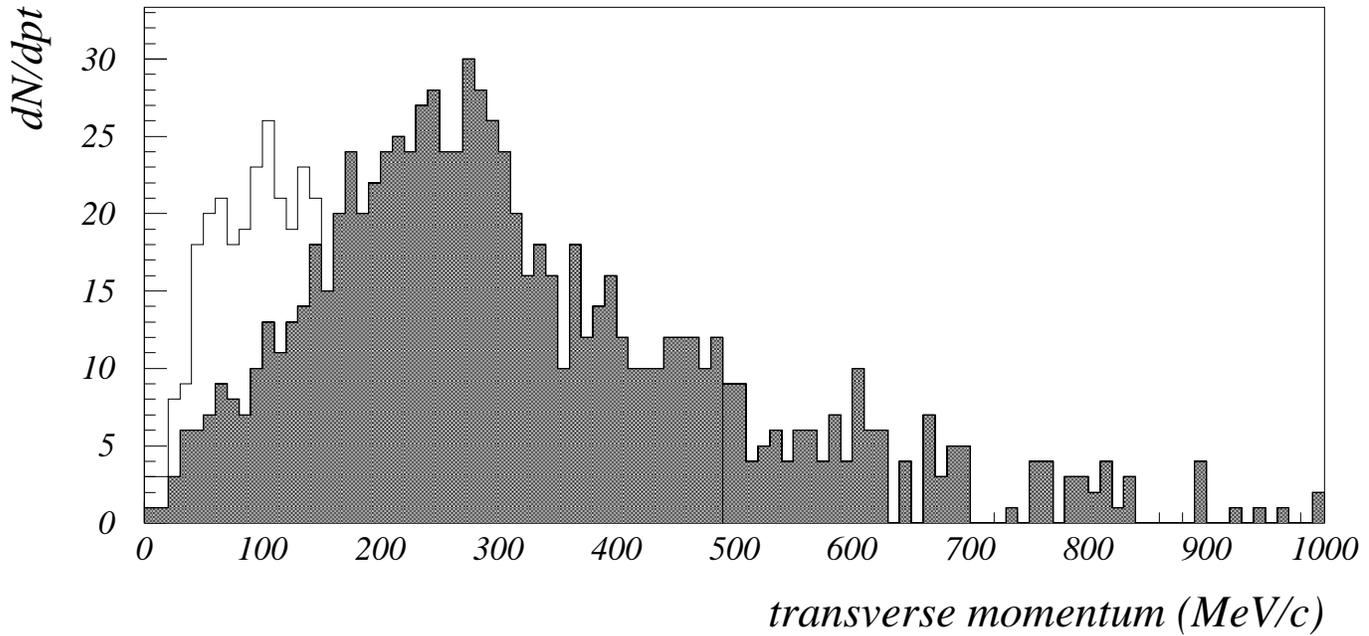
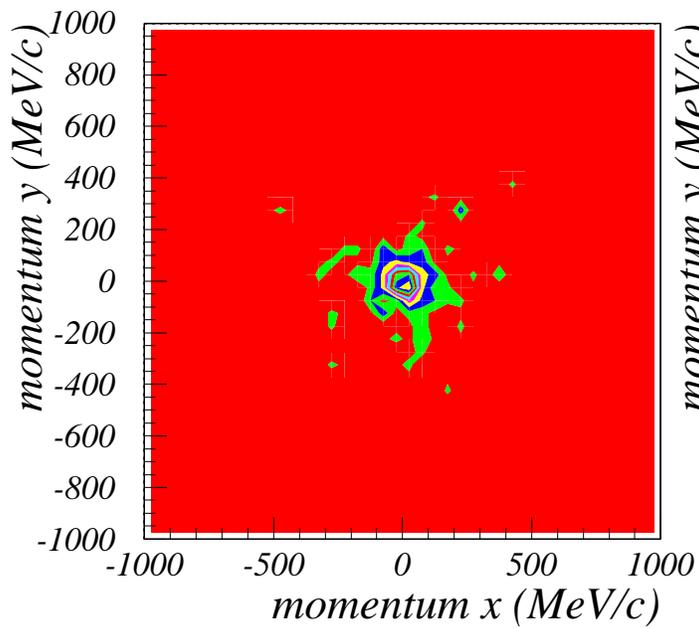
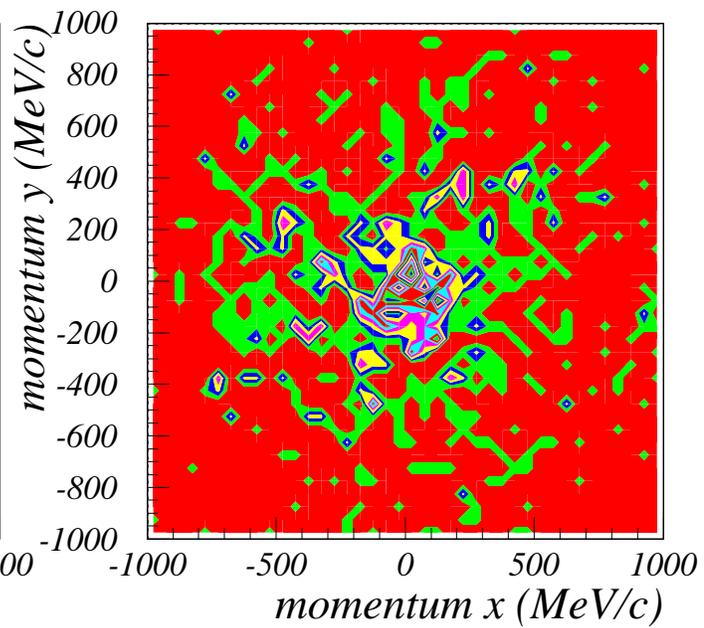
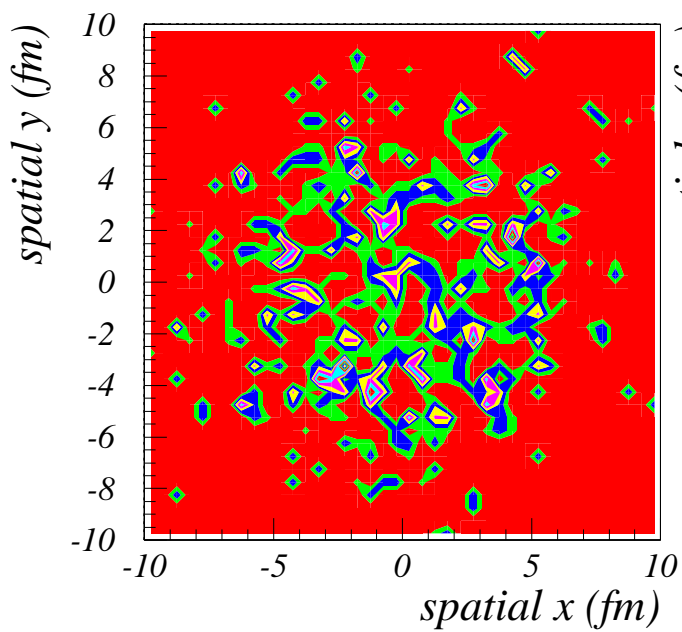
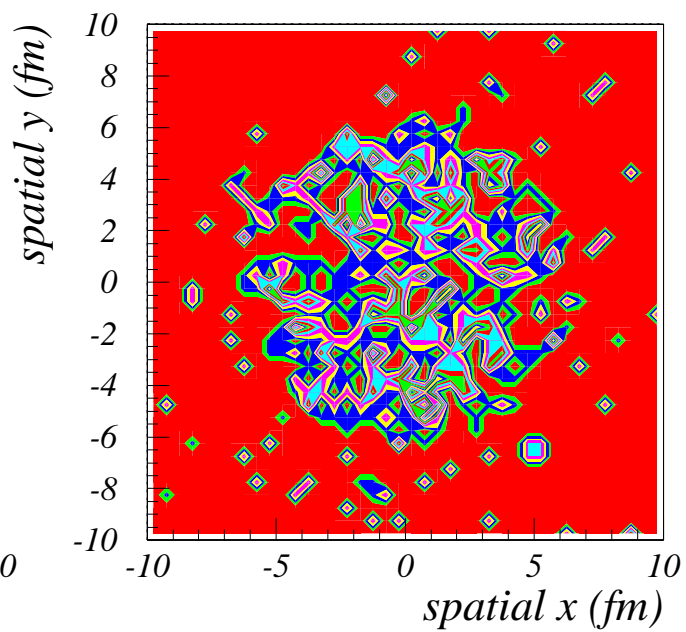
\end{document}